\documentclass[preprint,showpacs,preprintnumbers,amsmath,amssymb]{revtex4}

\usepackage{graphicx}% Include figure files
\usepackage{dcolumn}% Align table columns on decimal point
\usepackage{bm}% bold math

\begin{document}

\title{Primary population of antiprotonic helium states}

\author{J. R\'{e}vai}
\email{revai@rmki.kfki.hu}
 \affiliation{Research Institute for
Particle and Nuclear Physics, H-1525 Budapest, P.O.B. 49, Hungary}

\author{N. V. Shevchenko}
 \email{shev@thsun1.jinr.ru}
\affiliation{Joint Institute  for Nuclear Research, Dubna, 141980,
Russia}

\date{\today}

\begin{abstract}
A full quantum mechanical calculation of partial cross-sections
leading to different final states of antiprotonic helium atom was
performed. Calculations were carried out for a wide range of
antiprotonic helium states and incident (lab) energies of the
antiproton.
\end{abstract}

\maketitle

\section{Introduction}
One of the most impressive success stories of the last decade in
few-body physics is the high precision experimental and
theoretical studies of long lived states in antiprotonic helium
(for an overview see~\cite{yamazaki}). While the energy levels
have been both measured and
calculated to an extreme precision, allowing even for improvement
of numerical values of fundamental physical constants, some other
relevant properties of these states were studied with considerably
less accuracy. Among these is the formation probability of
different metastable states $(J,v)$ in the capture reaction
\begin{equation}
\label{react}
\bar p \, + \, ^4He \longrightarrow (^4He^+ \, \bar p)_{Jv} + e^-.
\end{equation}
The existing calculations of the capture rates of slow antiprotons in
$He$~\cite{koren1,koren2,cohen} are based on classical or semiclassical
approaches and they mainly
address the reproduction of the overall fraction (3\%)
of delayed annihilation events. Recent experimental results from
the ASACUSA project~\cite{ASACUSA}, however, yield some information on individual
populations of different metastable states, and our aim is to
perform a fully quantum mechanical calculation of the formation
probability of different states in the capture reaction.

\section{Calculation Method}
The exact solution of the quantum mechanical four-body problem,
underlying the reaction~(\ref{react}) is far beyond the scope of this work,
and probably also of presently available calculational
possibilities. Still, we want to make a full quantum mechanical,
though approximate, calculation of the above process. Full is
meant in the sense that all degrees of freedom are taken
explicitly into account, all the wave functions we use, are true
four-body states.

The simplest way to realize this idea is to use the plane wave
Born approximation which amounts to replacing in the
transition matrix element the exact scattering wave function
$\Psi_i^+$ by its initial state $\Phi_i$ which preceded the
collision.
\begin{equation}
\label{Tfi}
 T_{fi} = \langle \Phi_f | V_f | \Psi_i^+ \rangle \approx
 \langle \Phi_f | V_f | \Phi_i \rangle.
\end{equation}
In our case the initial and final wave functions were taken in the
form:
\begin{eqnarray}
\nonumber
 \Phi^i_{He,\, {\bf K}_i} ({\bf r}_1,{\bf r}_2,{\bf R}) &=&
 \Phi_{He}({\bf r}_1,{\bf r}_2) \,
 \frac{1}{(2 \pi)^{3/2}} \, e^{i {\bf K}_i {\bf R}} \\
\nonumber
 \Phi^f_{Jv,\, {\bf K}_f} (\mbox{\boldmath{$\rho$}}_1,
 \mbox{\boldmath{$\rho$}}_2,{\bf R}) &=&
 \Phi_{Jv}(\mbox{\boldmath{$\rho$}}_1,{\bf R}) \,
 \frac{1}{(2 \pi)^{3/2}} \, e^{i {\bf K}_f \mbox{\boldmath{$\rho$}}_2}
\end{eqnarray}
where ${\bf r}_i$ are the vectors pointing from helium to the
$i$-th electron, ${\bf R}$ is the vector between $He$ and
$\bar{p}$, while $\mbox{\boldmath{$\rho$}}_i$ are the Jacobian
vectors connecting the electrons with the center of mass of the
$He-\bar{p}$ system. For the $He$ the ground state wave function
we used the simplest effective charge hydrogen-like
ansatz~\cite{bethe}
\begin{equation}
\label{wfHe}
 \Phi_{He}(\bf{r}_1,\bf{r}_2) = N \, \exp{(-\sigma (r_1 + r_2))}.
\end{equation}

For the antiprotonic helium wave function we used the
Born-Oppenheimer form~\cite{shim,revai}, which correctly reflects
the main features of the final state:
\begin{equation}
\label{wfpHe}
 \Phi_{Jv}(\mbox{\boldmath{$\rho$}},{\bf R}) =
 \frac{\chi_{Jv}(R)}{R}\,
Y_{JM}(\hat{R}) \phi_{1\sigma}(\mbox{\boldmath{$\rho$}};{\bf R})
\end{equation}
where $\phi_{1\sigma}(\mbox{\boldmath{$\rho$}};{\bf R})$ is a
ground state two-center wave function, describing the electron
motion in the field of $He$ and $\bar{p}$ separated by a fixed
distance $R$, while $\chi_{Jv}(R)$ is the heavy-particle relative
motion wave function corresponding to $({}^4He \, \bar{p} \; e^-)$
angular momentum $J$ and ''vibrational'' quantum number $v$.
The transition potential $V_f$ is obviously
$$
 V_f = - \, \frac{2}{{\bf r}_2} + \frac{1}{|{\bf r}_1 - {\bf R}|} +
  \frac{1}{|{\bf r}_1 - {\bf r}_2|}_{\, .}
$$
The partial cross-section leading to a certain antiprotonic helium
state $(Jv)$ can be written as
\begin{equation}
\label{sig_int}
\sigma_{Jv} = (2\pi)^4 \,{K_f\over K_i}\, \mu_i \,\mu_f\int
d\Omega_{{\bf K}_f} \left|\langle\Phi^f_{Jv, {\bf K}_f}|V_f|\,
\Phi^i_{He, {\bf K}_i}\rangle \right|^{\, 2}_{\, .}
\end{equation}
The angular integrations occurring in the evaluation of Eq.~(\ref{sig_int})
were carried out exactly, using angular moment algebra, while the
3-fold radial integrals were calculated numerically.

The general expression~(\ref{sig_int}) for the cross-section leading to a
specific state $(Jv)$ can be rewritten in terms of matrix element
between angular momentum eigenstates as
\begin{equation}
\label{sig_sum}
 \sigma_{Jv} = (2\pi)^4 \,{K_f\over K_i}\, \mu_i \,\mu_f
 \sum_{\Lambda,l} (2 \Lambda + 1) \,
 |M_{J,l}^{\Lambda}|^2
\end{equation}
with
\begin{equation}
\label{me}
 M_{J,l}^{\Lambda}=
 \langle \, [\Phi_{Jv}
 \, \phi_{K_f,l}(\mbox{\boldmath{$\rho$}}_2)]_{M}^{\Lambda} \,  | \,
 V_f \, | \, [\Phi_{He} \, \phi_{K_i,\Lambda}({\bf
 R})]_{M}^{\Lambda} \, \rangle,
\end{equation}
where $\phi_{K,l}({\bf r})$ denotes free states which definite
angular momentum
$$
 \phi_{K,l}({\bf r}) = \sqrt{\frac{2}{\pi}} \, j_l(Kr) Y_{lm}(\hat{r})
$$
and $[\,\,\,]_M^J$ stands for vector coupling. Since the angular
momentum of the $He$ ground state is zero, the total angular
momentum $\Lambda$ of the incident side is carried by the
antiproton. A given antiprotonic helium final state $Jv$ can be
formed with different total angular momenta $\Lambda$ depending on
the orbital momentum $l$ carried away by the emitted electron. Our
calculations show, that only the values $l = 0$ and $l = 1$ give a
non-negligible contribution to the sum in Eq.~(\ref{sig_sum}).

\section{Results and Discussion}

We have calculated the partial population cross-sections
$\sigma_{Jv}$ for states with angular momentum $J$ and energy
$E_{Jv}$ in the interval $J=25-51$,\ $E_{Jv}=-(4.9-2.3)$ a.u.  For
these states all the energetically allowed transitions were
calculated for incident antiproton energies in the range $5-30$\
eV.

\begin{figure}
\includegraphics[scale=0.7]{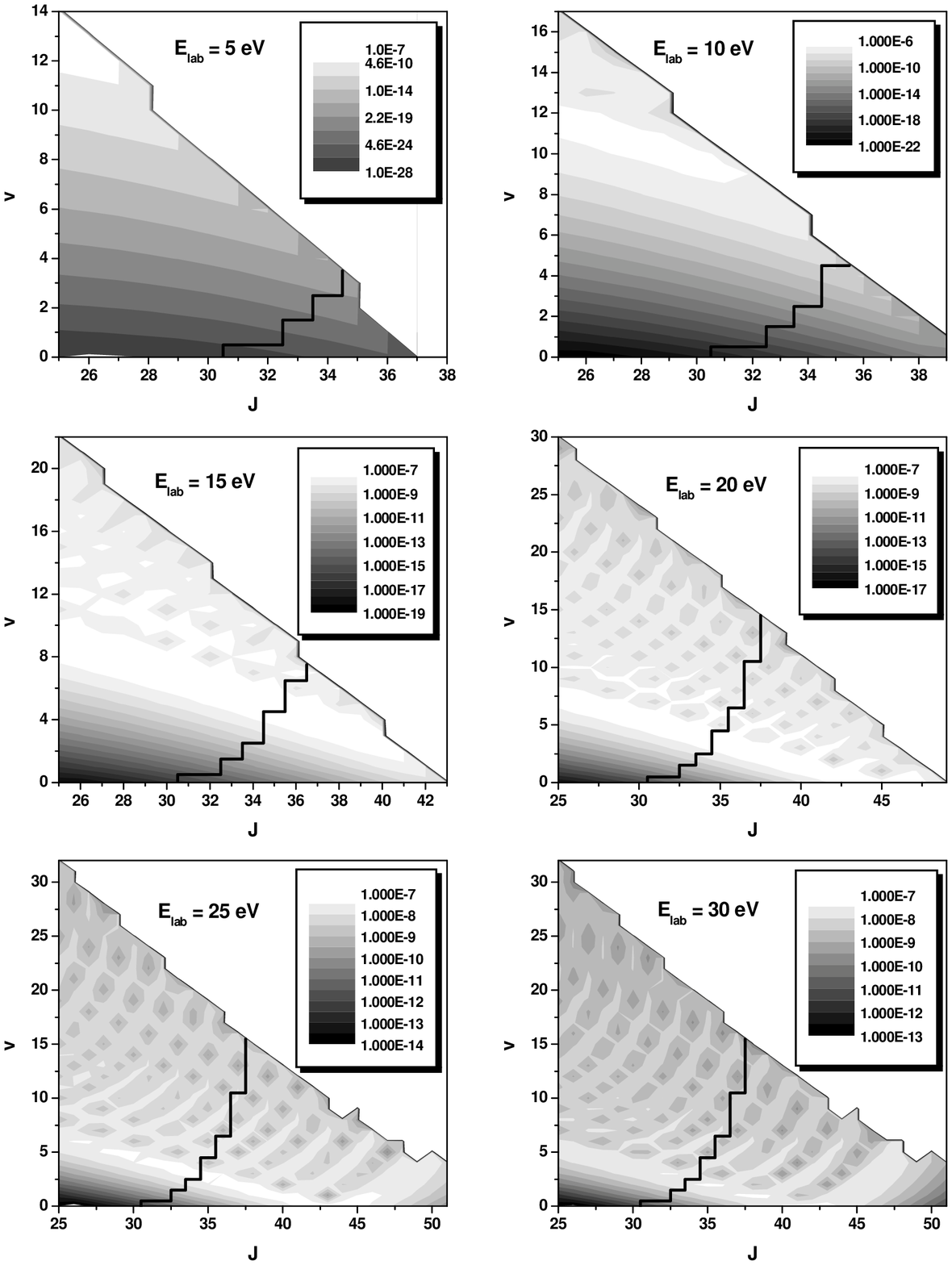}
\caption{\label{fig1.fig}
Overall distribution of calculated cross-sections over the
quantum numbers $J$ and $v$ for different incident antiproton
energies. The black line separates the short lived (on the left)
and long lived states (on the right side of the line) of antiprotonic
helium. All cross-sections are measured in units of $a_0^2$,
$a_0$ being the atomic length unit.
}\end{figure}

Our overall results are presented on the contour plots of Fig.~\ref{fig1.fig}.
The black line separates the regions of short-lived and long-lived
states. The latter (on the right side of the line) are selected
according to the usual criterium of Auger-electron orbital
momentum $l_{Auger}\ge 4$.

\begin{figure}
\includegraphics[scale=0.7]{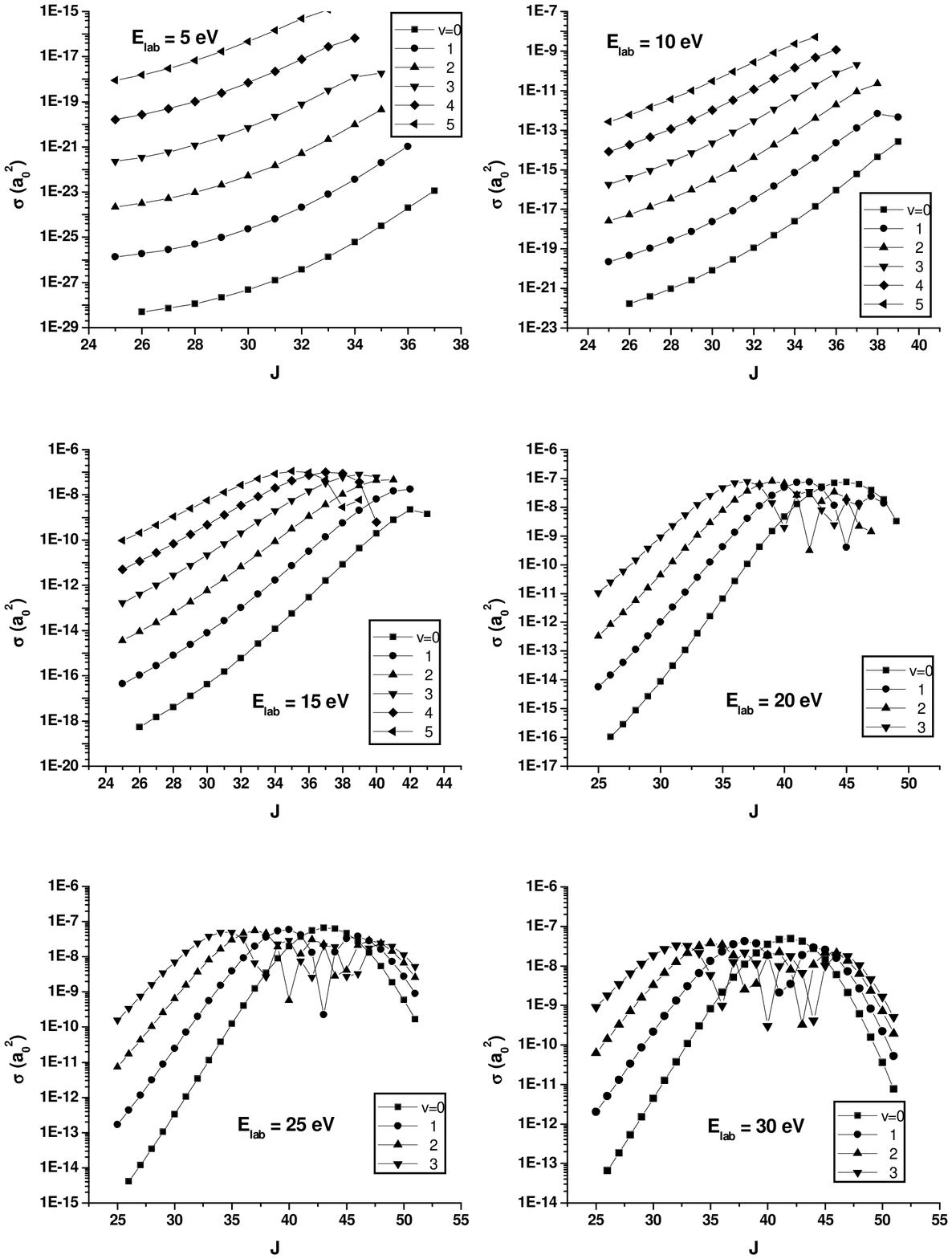}
\caption{\label{fig2.fig}
Cross-sections for the lowest few vibrational quantum numbers $v$ and
different incident antiproton energies.
}\end{figure}
\begin{figure}
\includegraphics[scale=0.6,angle=-90]{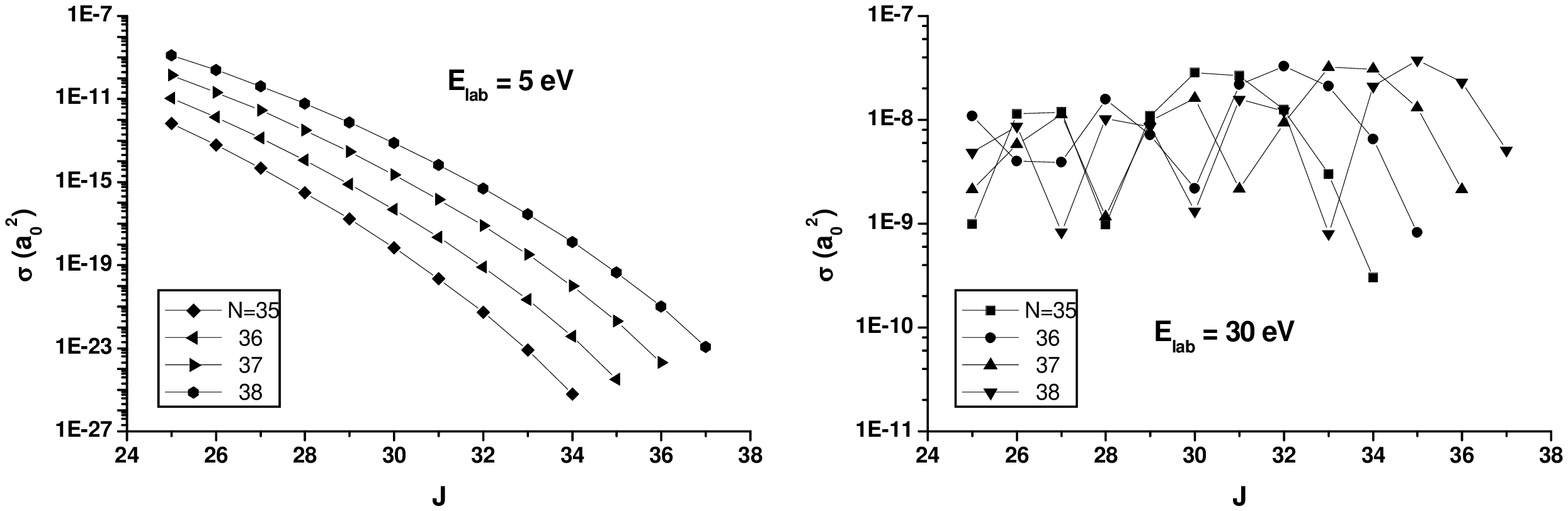}
\caption{\label{fig3.fig}
Examples of cross-sections with fixed $N$ in the "smooth" and
"oscillatory" regime (see text).
}\end{figure}
\begin{figure}
\includegraphics[scale=0.6,angle=-90]{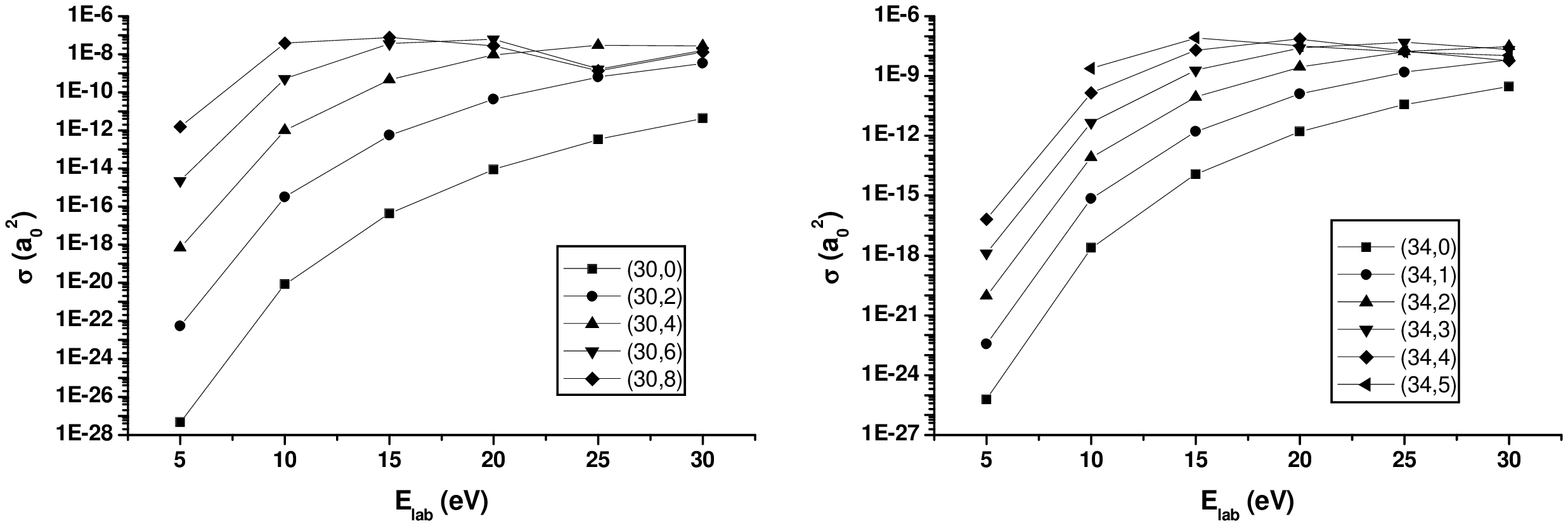}
\caption{\label{fig4.fig}
Dependence of selected cross-sections on incident antiproton energy.
}\end{figure}

In Figs.~\ref{fig2.fig}--\ref{fig4.fig} we tried to illustrate the dependence of
certain selected cross-sections $\sigma_{Jv}(E)$ on their parameters. In
Fig.~\ref{fig2.fig} we displayed certain cross-sections for various incident
energies as a function of antiprotonic helium angular momentum
$J$, connecting points, which correspond to a certain vibrational
quantum number $v$, while in Fig.~\ref{fig3.fig} the connected points belong to
the same principal quantum number $N=J+v+1$.

 Fig.~\ref{fig4.fig} shows the
dependence of a few cross-sections on the incident energy of
antiproton.

A table, containing all our results, would not fit the size of
this paper, however, it can be obtained from the authors upon
request, or seen/downloaded at
\url{http://www.rmki.kfki.hu/~revai/table.pdf}.

It is not easy to draw general conclusions about the relevant
physics of antiproton capture from this bulk of data. There is,
however, a conspicuous feature of our data, which deserve some
consideration. In a certain region of antiprotonic helium states
the dependence of the cross-sections on quantum numbers show a
smooth, regular pattern, while with increasing excitation energy
this behavior becomes irregular. On Fig.~\ref{fig1.fig} this is seen as a
transition from almost parallel stripes to an "archipelago" type
structure, while on Figs.~\ref{fig2.fig}--\ref{fig3.fig}
the smooth lines become oscillatory.

In order to reveal the origin of this phenomenon we looked into
the structure of the matrix element in Eq~(\ref{me}) and found that its
actual value is basically determined by the integration over $R$.
We can rewrite~(\ref{me}) as
\begin{equation}
\label{Fpot}
 M_{J,l}^{\Lambda} \sim\int_{0}^{\infty} \chi_{Jv}(R) \, F(R)
  \, j_{\Lambda}(K_i R)
 \, R \, dR,
\end{equation}
where $\chi_{Jv}(R)$ is the $He-\bar{p}$ relative motion wave
function in the BO state ${Jv}$, $j_{\Lambda}(K_i R)$ is the
spherical Bessel function of the incident $\bar{p}$, and $F(R)$
contains all rest: the potentials, the angular integrals, and the
integrals over the electron coordinates.
The expression (\ref{Fpot}) can be considered as a kind of ''radial,
one-dimensional'' Born approximation for the transition of the
antiproton from the initial state $j_{\Lambda}(K_i R)$ to the
final state $\chi_{Jv}(R)$ and $F(R)$ plays the role of the
potential.

\begin{figure}
\includegraphics[scale=0.7]{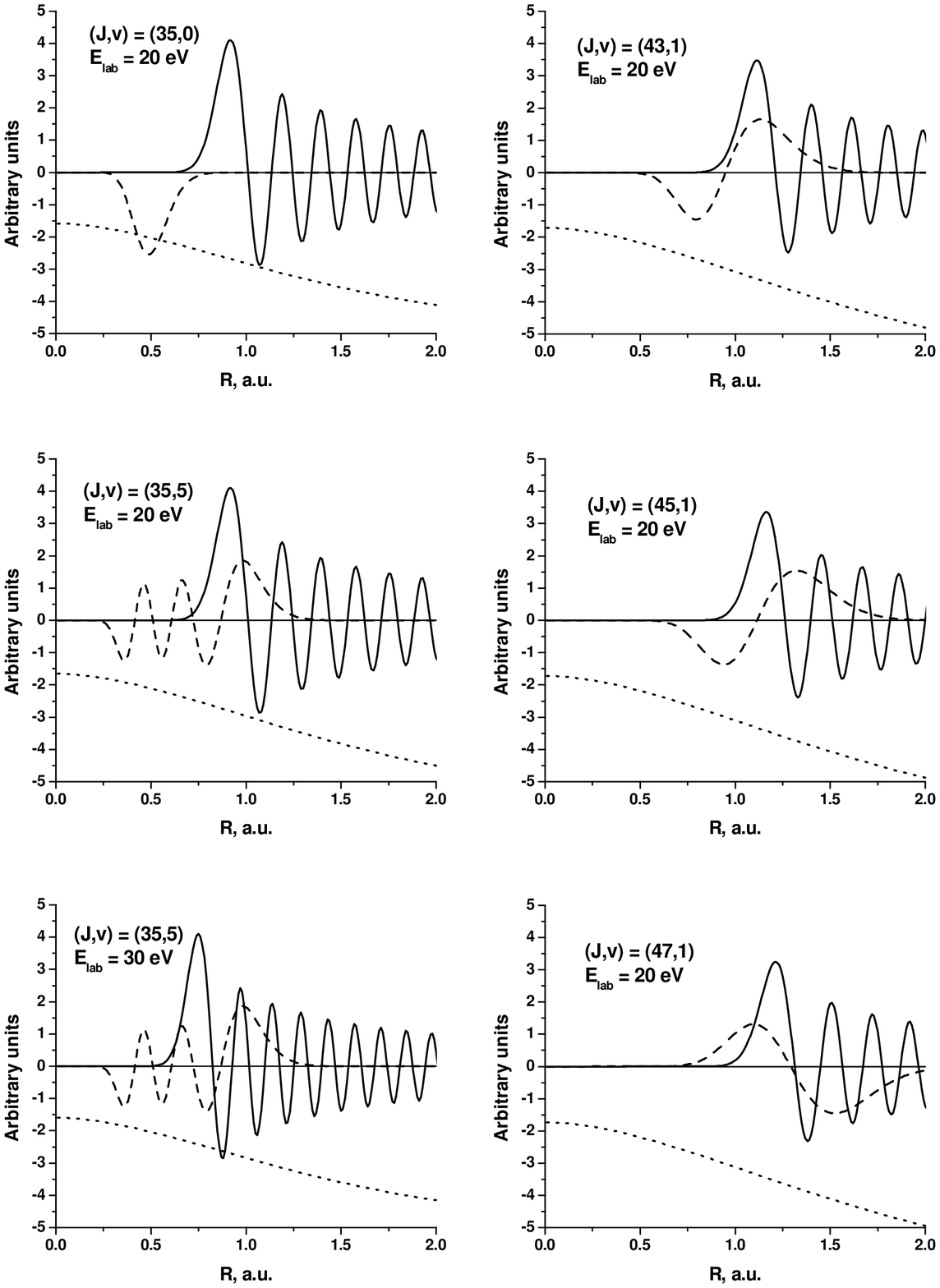}
\caption{\label{fig5.fig} Parts of the integrand of
Eq~(\ref{Fpot}): $j_{\Lambda}(K_iR)$ (solid line), $\chi_{Jv}(R)$
(dashed line) and $F(R)$ (dotted line). For details see text.
}\end{figure}

Fig.~\ref{fig5.fig} shows a few characteristic plots of $j_{\Lambda}(K_i R)$,
$\chi_{Jv}(R)$, and $F(R)$. It can be seen, that $F(R)$ depends
very weakly on the quantum numbers of the transition, thus its
interpretation as ''transition potential'' is not meaningless.
The other essential conclusion from Fig.~\ref{fig5.fig} is, that the value of
the integral is basically determined by the overlap of two rapidly
varying functions, $j_{\Lambda}(K_i R)$ and $\chi_{Jv}(R)$. While
$\chi_{Jv}(R)$ is strongly localized with rapid decay in both
directions $R \to \infty$ and $R \to 0$, $j_{\Lambda}(K_i R)$ is
rapidly oscillating for large $R$ and --- due to the high angular
momentum $\Lambda$ --- strongly decreasing in the direction $R \to
0$. For increasing $J$, $\chi_{Jv}$ slightly moves outwards, while
increasing $v$ (the number of its nodes) makes it more and more
oscillating. Increasing incident energy moves $j_{\Lambda}(K_i R)$
inwards. According to these general observations, the ''smooth
regime'' of the dependence of partial cross-sections on the energy
and quantum numbers, corresponds to the situation, when only the
''outer'' tail of $\chi_{Jv}(R)$ and the ''inner'' tail of
$j_{\Lambda}(K_i R)$ overlap. For increasing energy (incident or
excitation) the oscillating parts of $\chi_{Jv}$ and $j_{\Lambda}$
might overlap leading to an irregular ''unpredictable'' dependence
on quantum numbers and incident energy.

This idea can be  traced on the last two plots of Fig.~\ref{fig2.fig}.
The $\chi_{Jv}(R)$ functions for a given $v$ are essentially of the
same form, only with increasing $J$ they are pushed outwards into
the region of oscillations of $j_{\Lambda}(K_i R)$. For the
nodeless $v=0$ function this leads to a decrease of the integral
in Eq.~(\ref{Fpot}), while each node of the $v\ne 0$ functions produces a
minimum in the cross-section when it penetrates into the region of
non-vanishing $j_{\Lambda}(K_i R)$.

The three graphs on the right side of Fig.~\ref{fig5.fig}. demonstrate this
phenomenon on the case of the 20 eV $v=1$ curve of Fig.~\ref{fig2.fig}. It can
be seen, how the position of the node of $\chi_{Jv}(R)$ relative
to the first peak of the $j_{\Lambda}(K_i R)$ Bessel-function
brings about the minimum of the cross-section.

\section{Conclusions}
To our knowledge, this is the first calculation of the process~(\ref{react})
in which realistic final state wave functions were used. Due to
this fact we think, that our results concerning the relative
population of different final states might be reliable in spite of
the poor treatment of the dynamics of the capture process. As for
the absolute values of cross-sections, a more realistic dynamical
treatment of the reaction~(\ref{react}) is probably inevitable.

The transition matrix elements are basically determined by the
overlap of the BO function $\chi_{Jv}(R)$ and the incident Bessel
function $j_{\Lambda}(K_i R)$ of the antiproton. All the rest can
be incorporated into a potential-like function $F(R)$, which
weakly depends on the quantum numbers of the transition. This
feature will be probably preserved if a more realistic initial
state wave function (both for the electrons and the antiproton)
will be used.

The ''smooth'' regime of the quantum number dependence of the
partial cross-section allows to check the existing two ''thumb
rules'' \cite{yamazaki, shim} for the most likely populated
antiprotonic helium states. One of them states, that the mostly
populated levels will have
\begin{equation}
\label{thumb}
 N \sim \sqrt{\frac{M}{m}} \sim 37-38
\end{equation}
while according to the other assumption, the maximum of the
capture cross-section occurs for zero (or smallest possible)
energy of the emitted electron and correspondingly for highest
excitation energy. From our contour plots of Fig.~\ref{fig1.fig} we can
conclude, that the maximum cross-sections occur along a line,
which can be approximated by
\begin{equation}
\label{approx}
v(\sigma_{max})=a-b*J
\end{equation}
with $a\sim 15-20$ and $b \sim 0.4-0.45$, depending on incident
energy. This observation does not seem to confirm any of the
"thumb rules".

As for comparison of our results with the recently obtained
experimental data~\cite{ASACUSA}, we would like to make two
remarks. First, our calculations show, that the cross-sections
strongly depend on incident antiproton energy. Since the energy
distribution of the antiprotons before the capture is unknown, the
direct comparison with the observed data is impossible. Secondly,
any observed population data inevitably involve a certain time
delay after formation and thus the effect of ''depopulation'' due to
collisional quenching. Since this effect is absent from our
calculation, again, the comparison with experimental data is not
obvious.

\begin{acknowledgments}
One of the authors (JR) acknowledges the support from OTKA grants
T037991 and T042671, while (NVS) is grateful for the hospitality
extended to her in the Research Institute for Particle and Nuclear
Physics, where most of the work has been done. The authors wish to
thank A.T. Kruppa  for providing them with one of the necessary
computer codes.
\end{acknowledgments}

\bibliography{article}

\begin{thebibliography}{8}
\expandafter\ifx\csname natexlab\endcsname\relax\def\natexlab#1{#1}\fi
\expandafter\ifx\csname bibnamefont\endcsname\relax
  \def\bibnamefont#1{#1}\fi
\expandafter\ifx\csname bibfnamefont\endcsname\relax
  \def\bibfnamefont#1{#1}\fi
\expandafter\ifx\csname citenamefont\endcsname\relax
  \def\citenamefont#1{#1}\fi
\expandafter\ifx\csname url\endcsname\relax
  \def\url#1{\texttt{#1}}\fi
\expandafter\ifx\csname urlprefix\endcsname\relax\def\urlprefix{URL }\fi
\providecommand{\bibinfo}[2]{#2}
\providecommand{\eprint}[2][]{\url{#2}}

\bibitem[{\citenamefont{{Yamazaki {\it et al.}}}(2002)}]{yamazaki}
\bibinfo{author}{\bibfnamefont{T.}~\bibnamefont{{Yamazaki {\it et al.}}}},
  \bibinfo{journal}{Phys.\ Rep.} \textbf{\bibinfo{volume}{366}},
  \bibinfo{pages}{183} (\bibinfo{year}{2002}).

\bibitem[{\citenamefont{Korenman}(1996)}]{koren1}
\bibinfo{author}{\bibfnamefont{G.~Y.} \bibnamefont{Korenman}},
  \bibinfo{journal}{Hyperfine\ Interact.} \textbf{\bibinfo{volume}{101-102}},
  \bibinfo{pages}{81} (\bibinfo{year}{1996}).

\bibitem[{\citenamefont{Korenman}(2001)}]{koren2}
\bibinfo{author}{\bibfnamefont{G.~Y.} \bibnamefont{Korenman}},
  \bibinfo{journal}{Nucl.\ Phys.\ A} \textbf{\bibinfo{volume}{692}},
  \bibinfo{pages}{145c} (\bibinfo{year}{2001}).

\bibitem[{\citenamefont{Cohen}(2000)}]{cohen}
\bibinfo{author}{\bibfnamefont{J.~S.} \bibnamefont{Cohen}},
  \bibinfo{journal}{Phys.\ Rev.\ A} \textbf{\bibinfo{volume}{62}},
  \bibinfo{pages}{022512} (\bibinfo{year}{2000}).

\bibitem[{\citenamefont{{Hori {\it et al.}}}(2002)}]{ASACUSA}
\bibinfo{author}{\bibfnamefont{M.}~\bibnamefont{{Hori {\it et al.}}}},
  \bibinfo{journal}{Phys.\ Rev.\ Lett.} \textbf{\bibinfo{volume}{89}},
  \bibinfo{pages}{093401} (\bibinfo{year}{2002}).

\bibitem[{\citenamefont{Bethe and Salpeter}(1957)}]{bethe}
\bibinfo{author}{\bibfnamefont{H.~A.} \bibnamefont{Bethe}} \bibnamefont{and}
  \bibinfo{author}{\bibfnamefont{E.~E.} \bibnamefont{Salpeter}},
  \emph{\bibinfo{title}{Quantum mechanics of one- and two-electron atoms}}
  (\bibinfo{publisher}{Springer Verlag},
  \bibinfo{address}{Berlin-G{\"o}ttingen-Heidelberg}, \bibinfo{year}{1957}).

\bibitem[{\citenamefont{Shimamura}(1992)}]{shim}
\bibinfo{author}{\bibfnamefont{I.}~\bibnamefont{Shimamura}},
  \bibinfo{journal}{Phys.\ Rev.\ A} \textbf{\bibinfo{volume}{46}},
  \bibinfo{pages}{3776} (\bibinfo{year}{1992}).

\bibitem[{\citenamefont{R\'evai and Kruppa}(1998)}]{revai}
\bibinfo{author}{\bibfnamefont{J.}~\bibnamefont{R\'evai}} \bibnamefont{and}
  \bibinfo{author}{\bibfnamefont{A.~T.} \bibnamefont{Kruppa}},
  \bibinfo{journal}{Phys.\ Rev.\ A} \textbf{\bibinfo{volume}{57}},
  \bibinfo{pages}{174} (\bibinfo{year}{1998}).

\end{thebibliography}

\end{document}